\documentstyle[psfig]{mn}

\newif\ifAMStwofonts

\ifoldfss
  \ifCUPmtlplainloaded \else
    \NewTextAlphabet{textbfit} {cmbxti10} {}
    \NewTextAlphabet{textbfss} {cmssbx10} {}
    \NewMathAlphabet{mathbfit} {cmbxti10} {} 
    \NewMathAlphabet{mathbfss} {cmssbx10} {} 
  \fi
  \ifAMStwofonts
    \ifCUPmtlplainloaded \else
      \NewSymbolFont{upmath} {eurm10}
      \NewSymbolFont{AMSa} {msam10}
      \NewMathSymbol{\upi}     {0}{upmath}{19}
      \NewMathSymbol{\umu}     {0}{upmath}{16}
      \NewMathSymbol{\upartial}{0}{upmath}{40}
      \NewMathSymbol{\leqslant}{3}{AMSa}{36}
      \NewMathSymbol{\geqslant}{3}{AMSa}{3E}

    \fi
  \fi
\fi 

\ifnfssone
  \newmathalphabet{\mathit}
  \addtoversion{normal}{\mathit}{cmr}{m}{it}
  \addtoversion{bold}{\mathit}{cmr}{bx}{it}
  \newmathalphabet{\mathbfit} 
  \addtoversion{normal}{\mathbfit}{cmr}{bx}{it}
  \addtoversion{bold}{\mathbfit}{cmr}{bx}{it}
  \newmathalphabet{\mathbfss} 
  \addtoversion{normal}{\mathbfss}{cmss}{bx}{n}
  \addtoversion{bold}{\mathbfss}{cmss}{bx}{n}
  \ifAMStwofonts
    \ifCUPmtlplainloaded \else
      \UseAMStwoboldmath
      \makeatletter
      \new@mathgroup\upmath@group
      \define@mathgroup\mv@normal\upmath@group{eur}{m}{n}
      \define@mathgroup\mv@bold\upmath@group{eur}{b}{n}
      \edef\UPM{\hexnumber\upmath@group}
      \new@mathgroup\amsa@group
      \define@mathgroup\mv@normal\amsa@group{msa}{m}{n}
      \define@mathgroup\mv@bold\amsa@group{msa}{m}{n}
      \edef\AMSa{\hexnumber\amsa@group}
      \makeatother
      \mathchardef\upi="0\UPM19
      \mathchardef\umu="0\UPM16
      \mathchardef\upartial="0\UPM40
      \mathchardef\leqslant="3\AMSa36
      \mathchardef\geqslant="3\AMSa3E
    \fi
  \fi
\fi 

\ifnfsstwo
  \DeclareMathAlphabet{\mathbfit}{OT1}{cmr}{bx}{it}
  \SetMathAlphabet\mathbfit{bold}{OT1}{cmr}{bx}{it}
  \DeclareMathAlphabet{\mathbfss}{OT1}{cmss}{bx}{n}
  \SetMathAlphabet\mathbfss{bold}{OT1}{cmss}{bx}{n}
  \ifAMStwofonts
    \ifCUPmtlplainloaded \else
      \DeclareSymbolFont{UPM}{U}{eur}{m}{n}
      \SetSymbolFont{UPM}{bold}{U}{eur}{b}{n}
      \DeclareSymbolFont{AMSa}{U}{msa}{m}{n}
      \DeclareMathSymbol{\upi}{0}{UPM}{"19}
      \DeclareMathSymbol{\umu}{0}{UPM}{"16}
      \DeclareMathSymbol{\upartial}{0}{UPM}{"40}
      \DeclareMathSymbol{\leqslant}{3}{AMSa}{"36}
     \DeclareMathSymbol{\geqslant}{3}{AMSa}{"3E}
    \fi
  \fi
\fi 

\ifCUPmtlplainloaded \else
  \ifAMStwofonts \else 
    \def\upi{\pi}
    \def\umu{\mu}
    \def\upartial{\partial}
  \fi
\fi

\title[Ne and Ar emission lines in ionized gaseous nebulae]{Neon and Argon optical emission lines in ionized gaseous nebulae: Implications and applications}
\author[E. P{\'e}rez-Montero, G. F. H\"agele, T. Contini \& A.I. D{\'\i}az]
       {Enrique P{\'e}rez-Montero$^{1,2}$\thanks{Post-Doc fellow of Ministerio de Educaci\'on y
    Ciencia, Spain; enrique.perez@ast.obs-mip.fr}, Guillermo F. H\"agele$^{2}$\thanks{PhD fellow of Ministerio de Educaci\'on y Ciencia, Spain}, Thierry Contini$^{1}$ 
\newauthor and \'Angeles I. D\'{\i}az$^{2}$\thanks{On sabbatical leave at IoA, Cambridge}\\
$^{1}$ Laboratoire d'Astrophysique de Toulouse et Tarbes (UMR 5572). Observatoire Midi-Pyr\'en\'ees. 14, avenue Edouard Belin\\ F-31400. Toulouse. France\\ 
 $^{2}$ Departamento de F\'{\i}sica Te\'orica, C-XI, Universidad Aut\'onoma de
Madrid, 28049 Madrid, Spain\\  }
        
\date{Accepted 
      Received ;
      in original form April 2007}

\pagerange{\pageref{firstpage}--\pageref{lastpage}}
\pubyear{2006}

\begin{document}

\maketitle

\label{firstpage}

\begin{abstract}
In this work we present a study of the strong optical collisional emission lines of Ne and Ar in an heterogeneous sample of ionized gaseous nebulae for which it is possible to derive directly the electron temperature and hence the chemical abundances of neon and argon. We calculate using a grid of photoionization models new ionization correction factors for these two elements and we study the behaviour of Ne/O and Ar/O abundance ratios with metallicity. We find a constant value for Ne/O, while there seems to be some evidence for the existence of
negative radial gradients of  Ar/O over the disks of some nearby spirals. We study the relation between the intensities of the emission lines of [Ne{\sc iii}] at 3869 {\AA} and [O{\sc iii}] at 4959 {\AA} and 5007 {\AA}. This relation can be used in empirical calibrations and diagnostic ratios extending their applicability to bluer wavelengths and therefore to samples of
objects at higher redshifts. Finally, we propose a new diagnostic using [O{\sc ii}], [Ne{\sc iii}] and H$\delta$ emission lines to derive metallicities for galaxies at high z.

\end{abstract}


\section{Introduction}

The chemical history of the Universe can be investigated by studying the
behaviour of abundance ratios of different chemical
species as a function of metallicity, which is the main indicator of the
chemical evolution of a galaxy.  If two elements are produced by stars
of the same mass range, they will appear simultaneously in the
interstellar medium (ISM) and hence their relative abundance will be
constant. But if they are produced by stars of different mass ranges,
they will be ejected into the ISM in different time scales. 
The chemical abundances
of the elements heavier than hydrogen can be studied by measuring the fluxes of the absorption and emission
lines in the spectra of stars and galaxies. Unluckily, only the collisional emission lines emitted by the
ionized gas surrounding massive star clusters are detectable in most of galaxies. Since the brightest  emission
lines in the optical part of the spectrum are emitted by oxygen, this element has been taken as
the main tracer of metallicity for these objects. Nevertheless, the depletion of some of the most
important elements, including oxygen, onto dust grains, whose composition is difficult to ascertain, makes the determination
of these abundances more uncertain. This is not the case for the elements that
occupy the last group in the periodic table, that have an electronic configuration with the outer shell 
completely filled and they seldom associate to other elements and do not constitute part of the dust grains in the ISM.
Therefore, although the presence of these grains can affect the determination of metallicity in other ways, affecting the photoionization equilibrium of the gas, the uncertainty due to depletion factors has not to be considered in the determination of the chemical abundances of the
noble gases.

Neon and argon are products of the late stages in the evolution of massive stars. Neon is produced by carbon burning and 
is expected to track oxygen abundances very closely. The measurement of neon abundances in extragalactic HII regions (Garnett, 2002)
and planetary nebulae (Henry, 1989) confirm this trend, despite the uncertainties in the derivation of the ionization correction factor
of Ne. On the other hand, argon, like sulphur, is produced by oxygen burning and, again, it is expected to track O abundances. Nevertheless, as it is also the case for
sulphur,there is some evidence of decreasing values of Ar/O for higher metallicities (Garnett, 2002). The same trend has been observed for sulphur in halo metal poor stars (Israelian \& Rebolo, 2001) and extragalactic HII regions (D\'\i az et al., 1991; Garnett, 2002; P\'erez-Montero et al., 2006). This problem, perhaps, is related to the proximity of the production site of these elements to the stellar core and the yields would be sensitive to the conditions during the supernova explosion (Weaver \& Woosley, 1993). 

The emission lines of Ne and Ar are not as intense as some of the other strong lines
in the optical spectrum but there is a growing number of HII regions for which there are measurements with good signal-to-noise ratio.
This is the case of [Ne{\sc iii}] at 3869 {\AA}, whose blue wavelength, makes it observable in the optical spectrum of bright objects,
even at high redshifts, and the emission line of [Ar{\sc iii}] at 7136 {\AA}. The emission lines of these elements in the IR will
supply a great deal of worthy additional information but, at the moment, the objects with observations of the fine-structure lines of [Ne {\sc ii}], [Ne {\sc iii}], [Ar {\sc ii}], [Ar {\sc iii}] are still very scarce. The derivation of the chemical abundances of
these elements in HII regions can be calculated by means of the previous determination of some electron
collisional temperature and the previous knowledge of the ionization structure of the element whose abundance 
is required. The results of photoionization models point to a similar ionization structure of Ne and O on the one hand and 
S and Ar on the other ({\em e.g.} P\'erez-Montero \& D\'\i az, 2007). These facts make these lines suitable to be used as
substitutes of the bright emission lines of oxygen and sulphur in empirical indicators of metallicity ({\em e.g.} Nagao et al., 2006) or diagnostic ratios to distinguish starbursts galaxies from
active galactic nuclei ({\em e.g.} Rola et al., 1997).

In this paper, we investigate the element abundances of neon, argon and oxygen for a sample of objects that is described in Section 2. In Section 3, we derive
the physical conditions of these objects, including the calculation of the electron temperatures involved in the derivation of
Ne and Ar abundances. In Section 4, we describe the grid of photoionization models that we have used to derive a new
set of ionization correction factors for these two elements. These new ICFs are described in the next Section, along with the
discussion of the behaviour of Ne/O and Ar/O ratios and the use of the brightest emission lines of these two elements
as empirical calibrators of metallicity and diagnostic ratios. Finally, we summarize our results and we present our conclusions.

\section{Description of the sample}

The sample includes emission-line objects observed in the optical part of the
spectrum with the detection of, at least, one auroral line with high signal-to-noise ratio in
order to derive electron temperatures directly and, hence, ionic chemical abundances with less uncertainty. 
We have compiled for this sample the strong lines of [O{\sc ii}] at 3727 {\AA} and [O{\sc iii}] at 4959 {\AA}
and 5007 {\AA} in order to derive oxygen abundances.  We have compiled as well the [Ne{\sc iii}] emission line
at 3869 {\AA}, the [Ar{\sc iii}] emission line at 7136 {\AA} and the [Ar{\sc iv}] at 4740 {\AA} to measure the respective
ionic abundances of neon and argon. 
The compilation includes the objects used in P\'erez-Montero \& D\'\i az (2005), including HII regions in our Galaxy
and the Magellanic Clouds, Giant Extragalactic HII regions and HII Galaxies, with the addition of new objects which
are listed in Table \ref{refs}

This list includes 12 HII Galaxies from the Sloan Digital Sky Survey (SDSS\footnote{The SDSS site is at {\tt www.sdss.org}}) with very low metallicities  identified by
Kniazev et al. (2003) and whose spectra have been re-analysed to obtain the required line emission data. Among
these galaxies, 11 do not have observations of the [O{\sc ii}] emission line at 3727 {\AA}, which is the case as well for 183 SDSS galaxies of the sample compiled by Izotov et al. (2006).
The compilation has therefore 633 HII Galaxies, 176 GEHRs and 44 HII regions of the Galaxy and the Magellanic Clouds with a good determination of oxygen, neon and argon ionic abundances for those ions that have been observed.


\begin{table}
\begin{minipage}{85mm}
\vspace{-0.3cm}
\normalsize
\caption{Bibliographic references for the emission line fluxes of the compiled sample}
\begin{center}
\begin{tabular}{lcc}
\hline
\hline
Reference & Object type\footnote{GEHR denotes Giant Extragalactic HII Regions, 
HIIG, HII Galaxies and  DRH, Diffuse HII Regions} & Objects \\
\hline
Bresolin et al., 2004 & M51 GEHRs & 10 \\
Bresolin et al., 2005 & GEHRs & 32 \\
Bresolin, 2007 & M101 GEHRs & 3 \\
Crockett et al., 2006 & M33 GEHRs & 6 \\
Garnett et al., 2004 & M51 GEHRs & 2 \\
Guseva et al., 2003a & SBS 1129+576 & 2 \\
Guseva et al., 2003b & HS 1442+650 & 2 \\
Guseva et al., 2003c & SBS 1415+437 & 2 \\
H\"agele et al., 2006 & HIIG & 3 \\
H\"agele et al., 2007 & HIIG & 7 \\
Izotov et al., 1997 & IZw18  & 2 \\
Izotov \& Thuan, 1998 & SBS 0335-018 & 1 \\
Izotov et al., 1999 & HIIG & 3 \\
Izotov et al., 2001 & HIIG & 2 \\
Izotov et al., 2004 & HIIG & 3 \\
Izotov \& Thuan, 2004 & HIIG & 33 \\
Izotov et al. & SDSS galaxies & 309 \\
Kniazev et al. & SDSS galaxies & 11 \\
Lee et al., 2004 & KISS galaxies & 13 \\
Melbourne et al., 2004 & KISS galaxies & 12 \\
P\'erez-Montero \& D\'\i az, 2005 & All & 361 \\
Van Zee, 2000 & UGCA92 & 2 \\
Vermeij et al., 2002 & DRH & 5 \\

\hline
\end{tabular}
\end{center}
\label{refs}
\end{minipage}
\end{table}


The emission line intensities
have been taken directly from the literature once reddening corrected. The aperture effects of all the compiled
observations are negligible due to the compact nature of the observed objects and the high excitation state of
the more important involved emission lines.

\section{Physical conditions}

\subsection{Electron density and temperatures}

In order to derive oxygen, neon and argon abundances for the ions presenting 
the corresponding strong emission line we have selected objects allowing a direct determination of the electron temperature. All physical conditions have been derived from the appropriate ratios of emission lines and using
fittings to the values obtained with the task TEMDEN, in the case of electron densities and temperatures, and IONIC, in the case of ionic abundances of the software package IRAF \footnote{IRAF, Image Reduction and Analysis Facility, is distributed by the National Optical Astronomical Observatory} and which are based on the five-level statistical
equilibrium model (De Robertis, Dufour \& Hunt, 1987; Shaw \& Dufour, 1995). The references for the collision strengths of the ions involved
in our calculations are listed in Table \ref{at_data}.


\begin{table}
\begin{minipage}{85mm}
\vspace{-0.3cm}
\normalsize
\caption{Bibliographic references for the collision strengths of the forbidden lines for each ion.}
\begin{center}
\begin{tabular}{ll}
\hline
\hline
Ion & Reference \\
\hline
O{\sc ii} & Pradhan, 1976\\
O{\sc iii} & Lennon \& Burke, 1994 \\
Ne{\sc iii} & Butler \& Zeippen, 1994\\
S{\sc ii} & Ramsbottom, Bell \& Stafford, 1996\\
S{\sc iii} & Tayal \& Gupta, 1999\\
Ar{\sc iii} & Galavis, Mendoza \& Zeippen, 1995\\
Ar{\sc iv} & Zeippen, Le Bourlot \& Butler, 1987\\

\hline
\end{tabular}
\end{center}
\label{at_data}
\end{minipage}
\end{table}


Next, following the same procedure described in P\'erez-Montero \& D\'\i az (2003), we have used
for the calculation of the chemical abundance the electron temperature
associated with the zone where each ionic species stays. To do this we have applied different
expressions for the relations between the electron temperature of each zone.

Electron densities have been calculated for a subsample of 361 objects from the ratio of [S{\sc ii}] emission lines I(6717\AA)/I(6731\AA). It is
representative of the low excitation zone of the ionized gas, and it is therefore used for the calculation of electron temperatures
in this zone which depend on density. For the remaining the objects, for which no data about [S{\sc ii}] lines
exist or for which they give non valid values, we have considered a mean density of 50 particles per cm$^3$, according to
the low density values found in the rest of HII regions.

The electron temperature of O$^{2+}$, representative of the high excitation zone of the ionized gas, has been derived
from the ratio of [O{\sc iii}] emission lines (I(4959\AA)+I(5007))/I(4363\AA) for a subsample of 771 objects, while for the rest we have
used the relation with other measured electron temperatures in an inverse way as described below, 33 from t([S{\sc iii}]), 20 from t([O{\sc ii}]) and finally 4 from t([N{\sc ii}]). In this latter case, the objects are M51 GEHRs of high metallicity from Bresolin et al. (2004). We have
assummed as well the t([O{\sc iii}]) to be equal to t([Ne{\sc iii}]) and t([Ar{\sc iv}]) in order to calculate Ne$^{2+}$ and Ar$^{3+}$ abundances, respectively.

The electron temperature of O$^+$, representative of the low excitation zone of the ionized gas, has been derived
from the ratio of the [O{\sc ii}] emission lines  (I(3727\AA)/I(7319\AA)+I(7330\AA)) for a subsample of 311 objects. The auroral lines of [O{\sc ii}]
present higher uncertainties because of their lower signal-to-noise ratio and a higher dependence on reddening 
correction, due to their larger wavelength distance to the closest hydrogen recombination emission line.
Besides, they present a small contribution due to recombination, although in quantities not larger than
the usual reported errors. Additionally, this ratio has a strong dependence on electron density, which
makes the determination of this temperature and, hence, the determination of O$^+$ abundances very uncertain.
This can increase the uncertainty of total oxygen abundances, mostly in high metallicity - low excitation
HII regions (P\'erez-Montero \& D\'\i az, 2005). For the rest of the objects, we have used the grid of relations between t([O{\sc ii}]) and t([O{\sc iii}]) presented in P\'erez-Montero \& D\'\i az (2003), which takes into account the dependence of t([O{\sc ii}]) on density and its corresponding uncertainty. For the same 4 objects
of M51 for which neither t([O{\sc iii}]) nor t([SIII)]) have been measured we have taken t([N{\sc ii}]) as representative of the electron temperature of the low excitation zone using the ratio of [N{\sc ii}] emission lines (I(6548\AA)+I(6584\AA))/I(5755\AA).

In the case of the electron temperature of Ar$^{2+}$, we have assumed that the ionization structure of this ion is 
rather similar to that of S$^{2+}$, whose electron temperature has intermediate values between the high and low excitation
zones (Garnett, 1992). We have calculated directly the temperature of S$^{2+}$ from the ratio of [S{\sc iii}] emission lines  (I(9069\AA)+I(9532\AA))/I(6312\AA) for a subsample of 299 objects. For the rest of objects we have taken relations of this temperature with t([O{\sc iii}]). In the case of low excitation objects  (T$_e$ $<$ 10000 K), based on photoionization models (P\'erez-Montero \& D\'\i az, 2005):
\begin{equation}
t([S {\sc iii})] = 1.05 \cdot t[(O {\sc iii})] - 0.08
\end{equation}
\noindent while for the high excitation objects we have taken the empirical relation found by H\"agele et al. (2006) that accounts better for the high dispersion found in this relation for HII galaxies:
\begin{equation}
t([S{\sc iii}]) = (1.19\pm0.08) \cdot t([O{\sc iii}]) - (0.32\pm0.10)
\end{equation}

\subsection{Ionic abundances}

Each ionic abundance has been calculated using the most prominent emission lines for each ion and the appropriate
electron temperature. In the case of O$^+$, we have used the following expression using the intensity of the [O{\sc ii}] line
at 3727 {\AA}, the electron temperature t([O{\sc ii}]), or any other value associated to the low excitation zone and the electron density, $n$:
\begin{eqnarray} 
12+log\left(\frac{O^+}{H^+}\right) = \log\left(\frac{I(3727)}{I(H\beta)}\right)+5.992+\frac{1.583}{t}- \nonumber \\
-0.681 \cdot \log t +\log (1+0.00023 \cdot n )
\end{eqnarray}

In the case of 11 of the 12 very low metallicity HII Galaxies identified by Kniazev et al. (2003) and also for a subsample of 183 HII Galaxies compiled from Izotov et al. (2006), which have no data on the [O{\sc ii}] $\lambda$ 3272 \AA\ line, the O$^+$ abundances have been measured from the emission lines of [O{\sc ii}] at 7319 and 7330 \AA\ using the following expression:
\begin{eqnarray} 
12+log\left(\frac{O^+}{H^+}\right) = \log\left(\frac{I(7325)}{I(H\beta)}\right)+6.895+\frac{2.44}{t}- \nonumber \\
-0.58 \cdot \log t -\log (1+0.0047 \cdot n )
\end{eqnarray}

For O$^{2+}$ abundances we can use the [O{\sc iii}] emission lines at 4959 {\AA} and 5007 {\AA}, along with the
electron temperature of [O{\sc iii}]:
\begin{eqnarray} 
12+log\left(\frac{O^{2+}}{H^+}\right) =\log\left(\frac{I(4959)+I(5007)}{I(H\beta)}\right)+6.144+ \nonumber \\
+\frac{1.251}{t}-0.55 \cdot \log t 
\end{eqnarray}

These two ionic abundances allow to calculate the total abundance of oxygen in relation to hydrogen via
the expression:
\begin{equation}
\frac{O}{H} = \frac{O^++O^{2+}}{H^+}
\end{equation}

Abundances of Ne$^{2+}$ have been calculated from the [Ne{\sc iii}] emission line at 3869 {\AA} and the electron temperature of [O{\sc iii}] for a subsample of 773 objects:
\begin{eqnarray} 
12+log\left(\frac{Ne^{2+}}{H^+}\right) = \log\left(\frac{I(3869)}{I(H\beta)}\right)+6.486+ \nonumber \\
+\frac{1.558}{t}-0.504 \cdot \log t 
\end{eqnarray}

Finally, regarding Ar, we can measure Ar$^{2+}$, from the [Ar{\sc iii}] emission line at 7136 {\AA} and the electron
temperature of [S{\sc iii}] for a subsample of 572 objects using the following expression:
\begin{eqnarray} 
12+log\left(\frac{Ar^{2+}}{H^+}\right) =\ log\left(\frac{I(7136)}{I(H\beta)}\right)+6.157+ \nonumber \\
+\frac{0.808}{t}-0.508 \cdot \log t
\end{eqnarray}

 It is possible to measure as well the lines of [Ar{\sc iv}]
at 4713 and 4740 \AA. Nevertheless, the first one usually appears blended with a line of He{\sc i} at 4711 {\AA} that is
difficult to correct, so it is better to use the second one to calculate the ionic abundance of Ar$^{3+}$
through the electron temperature of [O{\sc iii}]:
\begin{eqnarray} 
12+log\left(\frac{Ar^{3+}}{H^+}\right) = \log\left(\frac{I(4740)}{I(H\beta)}\right)+4.705+ \nonumber \\
+\frac{1.246}{t}-0.156 \cdot \log t 
\end{eqnarray}

This abundance has been calculated for a subsample of 253 objects with a simultaneous measurement of the [Ar{\sc iii}] and [Ar{\sc iv}] 
emission lines.

\begin{figure}
\centering
\psfig{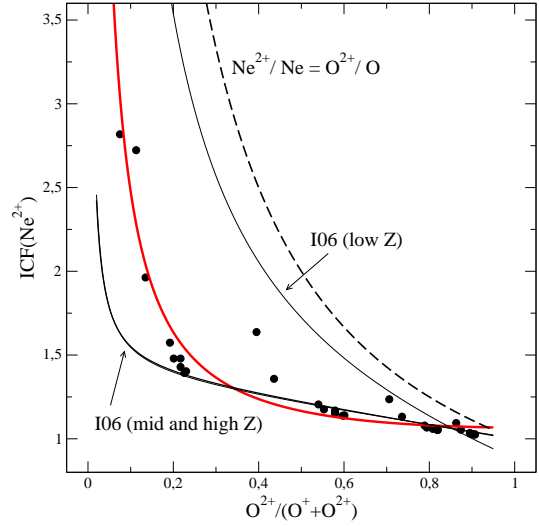}
\caption{Different relations between the ICF(Ne$^{2+}$) and the ratio of O$^{2+}$/O, approximated by O$^{2+}$/(O$^{+}$+O$^{2+}$). The points correspond to the models described in the text for stellar atmospheres of 45 kK and 50 kK. The thick solid line represents the best fit to these points. The thin solid lines correspond to different fits of photoionization models as a function of metallicity in Izotov et al. (2006). Finally the thick dashed line represents the classical approximation. }
\label{icf_ne}
\end{figure}

\section{Photoionization models}

A large grid of photoionization models has been calculated in order to check the validity
of the ionization correction factors for Ne and Ar.

We have used the photoionization code Cloudy 06.02 (Ferland et al., 1998), taking as ionizing
source the spectral energy distributions of O and B stars calculated with the code WM-Basic (version 2.11
\footnote{Available at http://www.usm.uni-muenchen.de/people/adi/Programs/Programs.html}, Pauldrach et al., 2001).
We have assummed spherical geometry, a constant density of 100 particles per cm$^{2}$ and a
standard fraction of dust grains in the interstellar medium. We have assummed as
well that the gas has the same metallicity as the ionizing source, covering the
values 0.05Z$_\odot$, 0.2Z$_\odot$, 0.4Z$_\odot$, Z$_\odot$ and
2Z$_\odot$, taking as the solar metallicity the oxygen abundance measured by
Allende-Prieto et al. (2001; 12+log(O/H) = 8.69). The rest of ionic abundances have been set
to their solar proportions.
A certain amount of depletion has been taken into account for the elements C, O, Mg, Si and Fe, considered
for the formation of dust grains in the code. Although it is expected that the dust-to-gas ratio scales with
metallicity (Shields \& Kennicutt, 1995), we have checked that different values of this ratio do not lead to
variations in our results about the computed ICFs in quantities larger than the reported errors.
Regarding other functional parameters we have considered different values of the ionization
parameter (log U = -3.5, -3.0, -2.5 and -2.0) and the stellar effective temperature (T$_*$ = 35000 K,
40000 K, 45000 K and 50000 K). This gives a total of 80 photoionization models to cover the
conditions of different ionized gas nebulae. Atomic data, including collision strengths 
in the models are consistent with those used in the calculation of chemical abundances. We have checked 
as well the influence of varying dielectronic recombination rate coefficients in the code according with the most
recent values (Badnell, 2006 and references therein), but we have not found any relevant variation in
our results concerning the calculation of ICFs.

\section{Discussion}

\subsection{ionization correction factors (ICF)}

\begin{figure*}
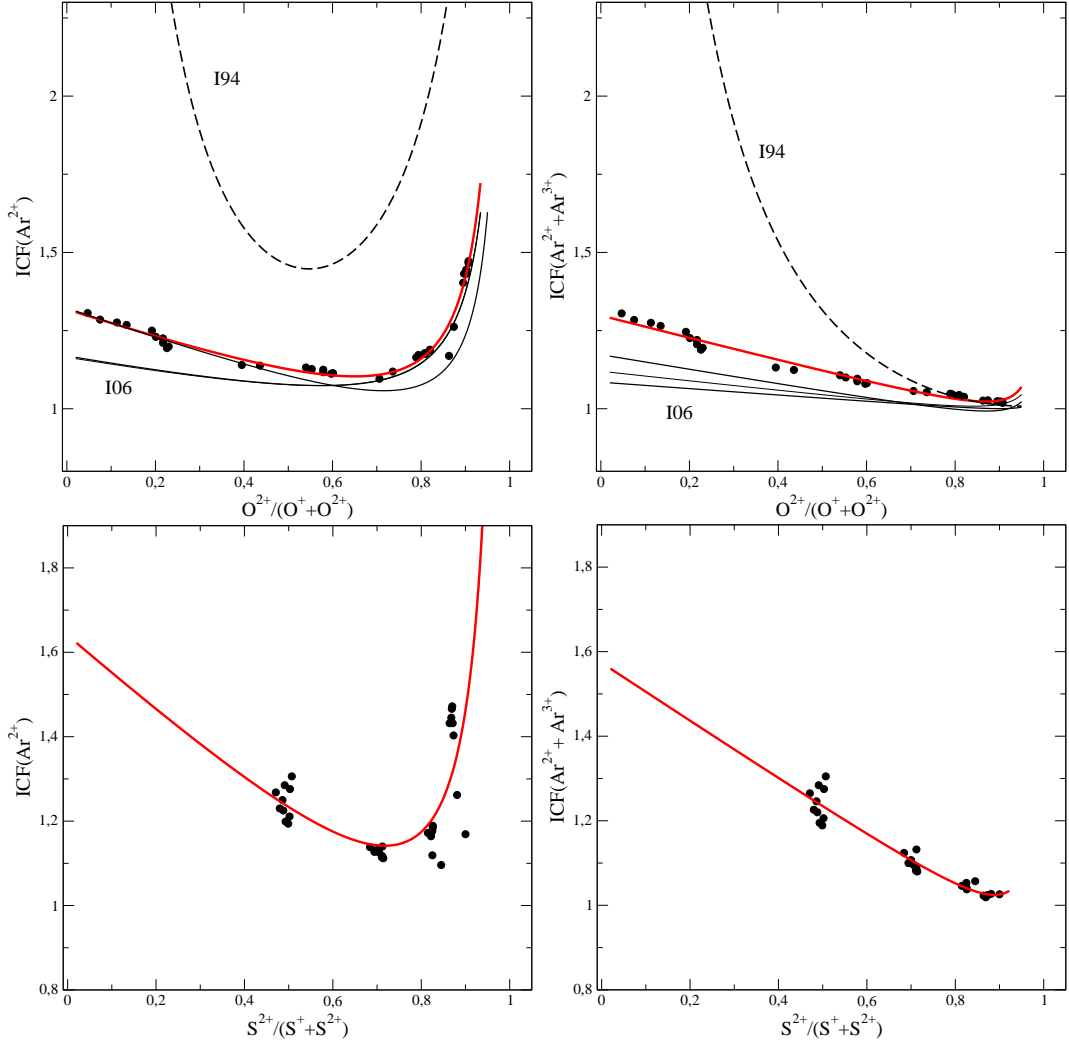

\begin{minipage}{170mm}
\centerline{
\psfig{figure=ICFArIII_OIII-ajuste-mas40kK.eps,width=7cm,clip=}
\psfig{figure=ICFArIII+ArIV_OIII-ajuste-mas40kK.eps,width=7cm,clip=}}
\centerline{
\psfig{figure=ICFArIII_SIII-ajuste-mas40kK.eps,width=7cm,clip=}
\psfig{figure=ICFArIII+ArIV_SIII-ajuste-mas40kK.eps,width=7cm,clip=}}
\caption{Representation of the ICF for Ar as a function of the O$^{2+}$/(O$^+$+O$^{2+}$) ratio in the two upper panels, and as a function of  S$^{2+}$/(S$^+$+S$^{2+}$) in the lower ones. Dashed lines represent the expressions given by Izotov et al. (1994), the thin solid lines represent those in Izotov et al. (2006). Finally, the points represent the models described here for cluster effective temperatures of 45 kK and 50 kK. The thick line is the best fit to these points.}
\label{icf_ar}
\end{minipage}
\end{figure*} 

ICFs stand for the unseen ionization stages of each element.

\begin{equation}
\frac{X}{H} = ICF(X^{+i}) \cdot \frac{X^{+i}}{H^+}
\end{equation}

where $X$ is the element whose ICF is required and $X^{+i}$ is the ionic species whose
abundance is calculated by means of the detected emission lines.

 In the case of neon,
only the [Ne{\sc iii}] line at 3869 {\AA} is detected in the optical spectrum in a large sample of objects. 
Since its ionization structure is quite similar to that of oxygen, the following approximation can be used:
\begin{equation}
\frac{Ne^{2+}}{Ne} \approx \frac{O^{2+}}{O}
\end{equation}

that leads to:
\begin{equation}
ICF(Ne^{2+}) = O/O^{2+} \approx (O^++O^{2+})/O^{2+}
\end{equation}

This relation is shown as a thick dashed line in Figure \ref{icf_ne}. 
Nevertheless, Izotov et al. (2004) point out from a set of photoionization models that this approximation
is less accurate in regions with a lower ionization parameter where the charge transfer reaction between
O$^{2+}$ and H$^0$ becomes more efficient. In this type of objects it is expected to find larger
abundances of Ne$^{2+}$ in relation to O$^{2+}$ and, hence, the O$^{2+}$/O ratio provides
larger ICFs.  This trend is confirmed with the set of photoionization models used by Izotov et al. (2006) that,
 besides, find a slight dependence of the ICF(Ne$^{2+}$) on metallicity, with larger values of this ICF for high metallicity regions.
This dependence is due to the addition of X-rays sources in their models with lower metallicity and not to a real dependence
on metallicity or ionization parameter.
The ICFs from Izotov et al. (2006) are shown as thin solid lines in Figure \ref{icf_ne}.
 The fit of our models for the hottest stars
(45000 K and 50000 K), whose effective temperatures reproduce better the radiation field in the most massive
ionizing cluster is shown as a thick solid line in Figure \ref{icf_ne}. It reveals the same overestimate of the classical approximation in low excitation objects. However, we do not find any noticeable dependence on
the metallicity. This fit gives:
\begin{equation}
iCF(Ne^{2+}) = 0.753 + 0.142\cdot x + 0.171/x 
\end{equation}

where x = O$^{2+}$ / (O$^{+}$+O$^{2+}$), with a rms of 0.074 only.

Regarding argon, the ICF can vary depending on the availibity of Ar$^{3+}$ abundances.
Izotov et al. (1994) propose the following formula to calculate the total abundance of Ar using the ionic abundances of these two ions:
\begin{eqnarray}
ICF(Ar^{2+}+Ar^{3+})\,=\,\Big[0.99+0.091\Big(\frac{O^+}{O}\Big) \nonumber\\ 
-1.14\Big(\frac{O^+}{O}\Big)^2+0.077\Big(\frac{O^+}{O}\Big)^3\Big]^{-1}
\end{eqnarray}

On the other hand, if only the emission line of [Ar{\sc iii}] is available,  they propose the following
expression:

\begin{equation}
ICF(Ar^{2+})\,=\,\Big[0.15+2.39\Big(\frac{O^+}{O}\Big)
-2.64\Big(\frac{O^+}{O}\Big)^2\Big]^{-1}
\end{equation}

Both formulae are shown as dashed lines in the two upper panels of Figure \ref{icf_ar}, respectively.
Nevertheless, these fits clearly overestimate the amount of total Ar compared with the expressions proposed by Izotov et al. (2006) who, 
as in the case of Ne, find a dependence with metallicity.
 These fits are shown in the upper panels of Figure \ref{icf_ar} as thin solid lines for low, intermediate and
high metallicities. Finally, the
fits to our own models are closer to these latter, but we do not find any relevant dependence of
our models on metallicity. The expressions for the fits of these models are, as a function of x = O$^{2+}$ / (O$^{+}$+O$^{2+}$):
\begin{equation}
ICF(Ar^{2+}+Ar^{3+}) = 0.928 + 0.364\cdot(1-x) + 0.006/(1-x) 
\end{equation}

\noindent with a rms of 0.011, shown as a thick solid line in the left upper panel of Figure \ref{icf_ar}, and
\begin{equation}
ICF(Ar^{2+}) = 0.596 + 0.967\cdot(1-x) + 0.077/(1-x) 
\end{equation}
\noindent with a rms of 0.067, shown as a thick solid line in the right upper panel of Figure \ref{icf_ar}. It is possible, as well, to express these ICFs as a function of sulphur abundances. This allows to derive total argon abundances using red or near-IR observations only, in the case of ICF of Ar$^{2+}$. Besides, these ICFs can be used in the subsample of SDSS objects whose [O{\sc ii}] line at 3727 {\AA} is not detected but which, on the contrary, have good detection of the [S{\sc iii}] at 9069 \AA. The fits of our models, as a function of x = S$^{2+}$ / (S$^{+}$+S$^{2+}$) are:
\begin{equation}
ICF(Ar^{2+}+Ar^{3+}) = 0.870 + 0.695\cdot (1-x) + 0.0086/(1-x) 
\end{equation}
\noindent with a rms of 0.019, which is shownd as a thick solid line in the left lower panel of Figure \ref{icf_ar}, and
\begin{equation}
ICF(Ar^{2+}) = 0.596 + 0.967\cdot (1-x) + 0.077/(1-x) 
\end{equation}
\noindent with a rms of 0.068, shown as a thick solid line in the right lower panel of Figure \ref{icf_ar}.

\subsection{Behaviour of Ne/O and Ar/O with metallicity}


\begin{figure}
\begin{minipage}{85mm}
\psfig{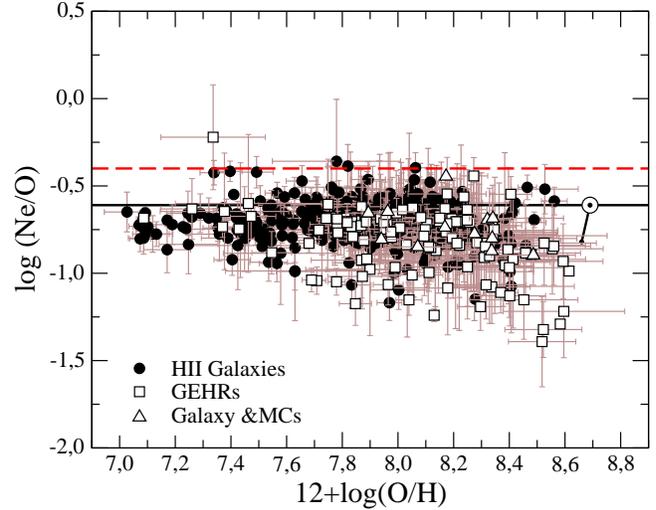}
\caption{Relation between the abundance ratio Ne/O and the metallicity for the sample of objects described in the text. Solid circles represent HII galaxies, open squares Giant Extragalactic HII regions and open triangles HII regions in the Galaxy and the Magellanic Clouds. We show as well the solar chemical abundances for O (Allende-Prieto et al. 2001) and for Ne (Grevesse \& Sauval, 1998), linked by a solid line with the solar Ne/O ratio from Asplund et al. (2005). The dashed line represents the ratio calculated by Drake \& Testa (2005) for a set of stellar coronae.}
\label{o_neo}
\end{minipage}
\end{figure} 

We have applied the ICF(Ne$^{2+}$) derived from our set of photoionization models to a total sample of 578 HII Galaxies, 117 GEHRs and
12 HII regions in our Galaxy and the Magellanic Clouds. In Figure \ref{o_neo}, we represent the ratio of Ne/O as a function of the total oxygen abundance, and we compare the obtained values with the solar one, taking the solar oxygen abundance to be 12+log(O/H) = 8.69 (Allende-Prieto et al., 2001) and the solar neon abundance to be 12+log(Ne/H) = 8.08 (Grevesse \& Sauval, 1998).  Although there is
a high dispersion, probably due to chemical inhomogeneities and different observational conditions, the values agree quite well with the assumption of a constant value of Ne/O, at least for low and intermediate metallicities. At higher metallicities, there is a slight decrease of Ne/O with O/H, perhaps due to the underestimate of the corresponding ICF in this regime, as pointed out by Izotov et al. (2006). Nevertheless, since we have not used additional X-ray sources in our models, this problem has to be further investigated. At this point,  it is necessary to stress that all the results from photoionization models are in contradiction with Vermeij \& Van der Hulst (2002) who find  
Ne$^+$ abundances even larger that those predicted by the classical approximation in a set of HII regions in the Galaxy and the Magellanic Clouds from direct ISO observations of the 
emission lines of [Ne{\sc ii}] and [Ne{\sc iii}] in the mid-infrared. Regarding the constant value of Ne/O, the average value of the sample, which is -0.72 $\pm$ 0.13,  is lower, although within the error, than our assumed solar one, but higher than the value reported by Asplund et al. (2005), connected to the value assumed here with a solid line in Figure \ref{o_neo} (log (Ne/O) = -0.82 in the photosphere). Nevertheless, all the derived values are far from the value found by Drake \& Testa (2005) using Chandra X-ray spectra in the coronae of 21 stars, which is log(Ne/O) $\approx$ -0.4. If this value is correct, then there is a clear underestimate of neon abundances as derived from optical collisional lines.

\begin{figure}
\begin{minipage}{85mm}
\psfig{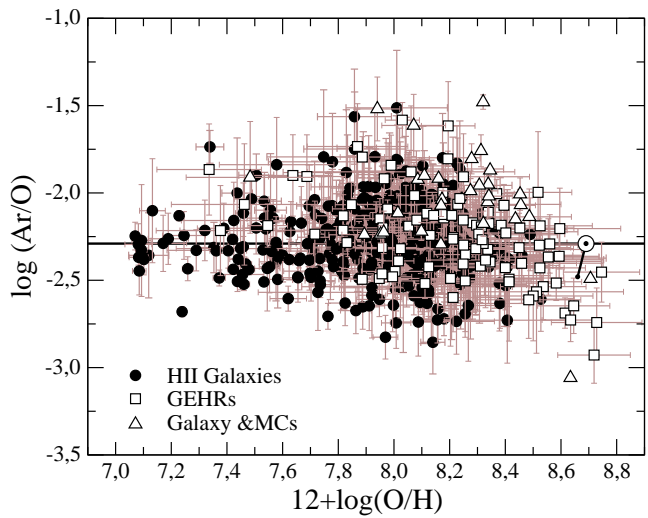}
\caption{Relation between the abundance ratio  Ar/O and the metallicity for the sample of objects described in the text. We show as well the solar chemical abundances for O (Allende-Prieto et al., 2001) and for Ar (Grevesse \& Sauval, 1998) linked by a solid line with the solar Ar/O ratio from Asplund et al. (2005).}
\label{o_aro}
\end{minipage}
\end{figure} 

\begin{figure}
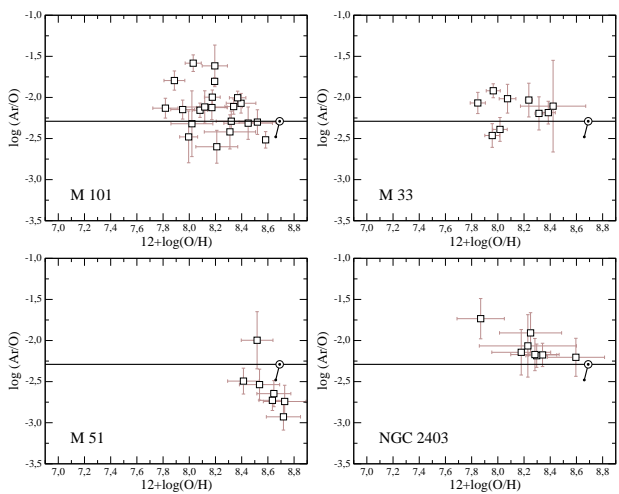

\centerline{
\psfig{figure=M101_aro.eps,width=4cm,clip=}
\psfig{figure=M33_aro.eps,width=4cm,clip=}}
\centerline{
\psfig{figure=M51_aro.eps,width=4cm,clip=}
\psfig{figure=N2403_aro.eps,width=4cm,clip=}}
\caption{Relation between the abundance ratio Ar/O and the metallicity for the disks of some spiral galaxies. From left to right and up to bottom: M101, M51, M33 and NGC 2403.}
\label{gradients_aro}

\end{figure} 

Regarding the relation between the Ar/O ratio and oxygen abundance, our results are shown in Figure \ref{o_aro}. In order to calculate total argon chemical abundances we have used ICFs based on the O$^{2+}$/O ratio for (Ar$^{2+}$ + Ar$^{3+}$) (104 objects) and Ar$^{2+}$ (344 objects), and ICFs based on the S$^{2+}$/(S$^{+}$+S$^{2+}$) ratio for the SDSS galaxies from Izotov et al.(2006) which do not have observation of the [O{\sc ii}] 3727 \AA\ line but for which measurements of the [S{\sc iii}] 9069 \AA\ line exist: 44 galaxies in the case of (Ar$^{2+}$ + Ar$^{3+}$) and 101 in the case of Ar$^{2+}$.
The value of Ar/O presents a larger dispersion than in the case of Ne/O, with an average value similar to the solar value from Allende-Prieto et al. (2001) and Grevesse \& Sauval (1998), represented by the solar symbol in Figure \ref{o_aro} and slightly higher than the value from Asplund et al. (2005), connected to it with a solid line. The relation appears to be slightly different for HII galaxies, whose behaviour is quite similar to that in the Ne/O diagram, and for Giant Extragalactic HII regions, for which there exists a slight trend of decreasing Ar/O for increasing metallicites. A more careful analysis for individual disk galaxies, as can be seen in Figure \ref{gradients_aro}, shows this trend more clearly,
contrary to it is expected for the production of an alpha element like Ar. This is compatible with the existence of larger Ar/O at lower effective radius, since there is evidence of negative gradients of metallicity in all these galaxies (M101: Kennicutt \& Garnett, 1996; M51: D\'\i az et al., 1991; M33 and NGC2403: Garnett et al., 1997). This result agrees quite well with that obtained for sulphur, both for Extragalactic HII regions (D\'\i az et al., 1991; P\'erez-Montero et al., 2006) and halo metal-poor stars (Israelian \& Rebolo, 2001).

\subsection{Empirical parameters based on Ne and Ar lines}

Although the [Ne{\sc iii}] emission line at 3869 {\AA} and the [Ar{\sc iii}] emission line at 7136 {\AA} are 
fainter than the oxygen emission lines, which are commonly used in different 
empirical calibrations of chemical abundances, they can be useful to ascertain metallicities when these lines are not available, either because they shift out of the optical region due to the object redshift or because the instrumental configuration does not cover the blue-green region of the spectrum.

The empirical calibrators are commonly used in objects whose low signal-to-noise and/or high metallicities do
not allow the accurate measurement of any of the auroral lines and, therefore, it is not possible to
derive the electron temperature and the ionic chemical abundances from the strong collisional lines.

\begin{figure}
\begin{minipage}{85mm}
\centerline{
\psfig{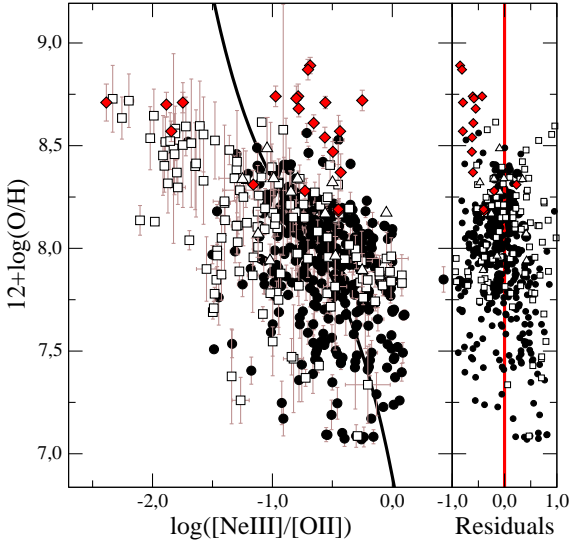}}
\caption{Relation between the logarithm of the ratio of the [Ne{\sc iii}] 3869 {\AA} and [O{\sc ii}] 3727 {\AA} lines and the oxygen abundance for the sample described in Section 2, along with a subsample of objects in our Galaxy and the Magellanic Clouds with oxygen abundance determinations from oxygen recombination lines (solid diamonds). In solid line, it is shown the relation proposed by Nagao et al. (2006) and in the right panel the residuals between the directly derived abundances and those obtained from this calibration.}
\label{o2_ne3}
\end{minipage}
\end{figure}

This is the case for the emission line ratio I([Ne{\sc iii}] 3869\AA)/I([O{\sc ii}] 3727\AA), proposed by Nagao et al. (2006) as
an empirical calibrator useful for high-redshift galaxies (up to z=1.6 in the optical part of the spectrum) and
relatively independent of reddening due to the proximity of the two lines. We show in Figure \ref{o2_ne3} the relation between
this ratio and the oxygen abundance for the sample of objects described in Section 2. This Figure shows a very high dispersion in
all the metallicity range but which is especially large in the high metallicity regime. This is confirmed
by the objects in the Galaxy and the Magellanic Clouds compiled by Peimbert et al. (2006) for which it exists
a direct determination of the oxygen abundance by means of recombination lines.
The residuals between the direct determination of
the oxygen abundance and those obtained from the relation based on the [Ne{\sc iii}]/[O{\sc ii}] ratio are shown in the right panel of the same
Figure. The standard deviation of these residuals in the whole metallicity regime reaches 0.83 dex.

\begin{figure}
\begin{minipage}{85mm}
 \centerline{
\psfig{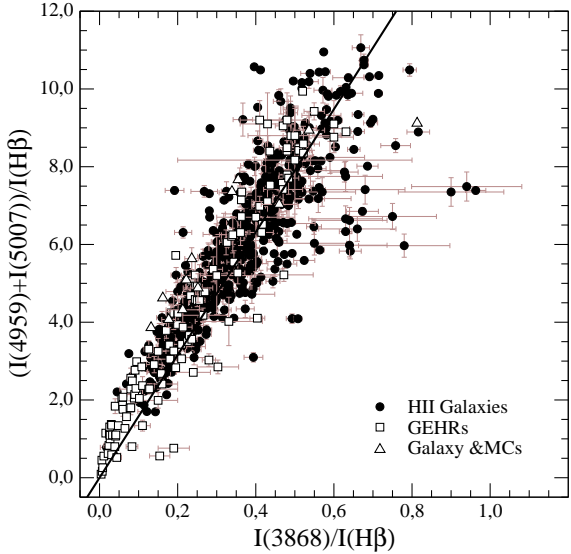}}
\caption{Relation between the intensity of the emission line of [Ne{\sc iii}] at 3869 {\AA} and the sum of the lines of [O{\sc iii}] at 4959 and 5007 {\AA} for the sample of objects described in Section 2. The solid line represents the best linear fit to the sample.}
\label{o3_ne3}
\end{minipage}
\end{figure} 

The reason for this huge dispersion can be found in the high dependence of the [Ne{\sc iii}]/[O {\sc ii}] ratio on ionization parameter, as it is
stressed by the same authors that propose it. This dependence is due to the tight relation between the flux of [Ne{\sc iii}] and [O{\sc iii}] emission
lines (see Figure \ref{o3_ne3}), as a consequence of the quite similar ionization structure of O$^{2+}$ and Ne$^{2+}$ and
the constant value of the Ne/O ratio.  The best linear fit between the fluxes of the lines for this sample yields:

\begin{equation}
I([O{\sc iii}] 4959+5007\AA)=(15.37 \pm 0.25) \cdot I([Ne{\sc iii}] 3869 \AA)
\end{equation}

Although, there is a deviation from this trend for some HII galaxies showing abnormally high values of the line of [Ne{\sc iii}] ({\em e.g.} UM382; Terlevich et al., 1991, HS1440+4302, HS1347+3811; Popescu \& Hopp, 2000), most of the objects are quite close to this relation. Even for the objects with low intensities, which correspond to high metallicity/low excitation regions and for which the charge transfer reaction between O$^{2+}$ and H$^0$ becomes more important, there exists a very good agreement with this linear fit. 
A possible cause to the deviation from this relation in some HII galaxies comes from the fact that Ne$^+$ has an ionization potential 5.85 eV larger than O$^+$ and therefore is more sensitive to the stellar effective temperature, which is higher in these HII galaxies.
One of the consequences of this relation is that the [Ne{\sc iii}]/[O{\sc ii}] ratio is, in fact, almost equivalent to [O{\sc iii}]/[O{\sc ii}] which is highly dependent on ionization parameter and effective temperature (P\'erez-Montero \& D\'\i az, 2005) and presents a lower dependence on metallicity.

The relation between [O{\sc iii}] and [Ne{\sc iii}] can be used as well in all the other diagnostic ratios involving [O{\sc iii}] lines in a similar way. This is the case of the O$_{23}$ parameter, also known as R$_{23}$, defined by Pagel et al. (1979) as the relative sum of the strong lines of [O{\sc ii}] at 3727 {\AA} and [O{\sc iii}] at 4959 and 5007 {\AA} in relation to H$\beta$. The relation between this parameter and the oxygen abundance is widely used for objects at large redshifts due to the relatively blue wavelength and high intensities of the involved lines. There exist many different calibrations of this parameter, whose main drawbacks are well known: (i) its relation with metallicity is double-valued with increasing values of O$_{23}$ for increasing values of metallicity in the low metallicity regime and decreasing values of O$_{23}$ for increasing metallicity in the high metallicity regime, requiring external methods to distinguish the upper from the lower branch and with a high dispersion in the middle-range, that makes  metallicites quite uncertain in the range 8.0 $<$ 12+log(O/H) $<$ 8.4; (ii) the dependence of O$_{23}$ on other functional parameters like ionization parameter or effective temperature. This is solved with the calibration of the parameter as a function of other quantities that reduce this dependence, as in the case of the [O{\sc ii}]/[O{\sc iii}] ratio (Kobulnicky et al., 1999 from the models of McGaugh, 1991) or the P parameter (Pilyugin, 2000); (iii) the lack of objects with directly derived abundances in the high metallicity regime makes the upper branch calibration to rely heavily on photoionization model results, which differ appreciably depending on the model
atmospheres used (Morisset et al. 2004) and the chosen input conditions. Besides, these models predict higher metallicities than those derived from electron temperatures of [N{\sc ii}] or [S{\sc iii}] in inner disk regions ({\em e.g.} Bresolin, 2006), but not with those derived from oxygen recombination lines. The difficulties found to derive accurate metallicites using this calibration and, in general, all the other empirical relations, make them appropriate to study distributions of metallicity in a statistical way in large surveys of emission lines objects, but they do not offer reliable results for individual determinations.

\begin{figure}
\begin{minipage}{85mm}
\psfig{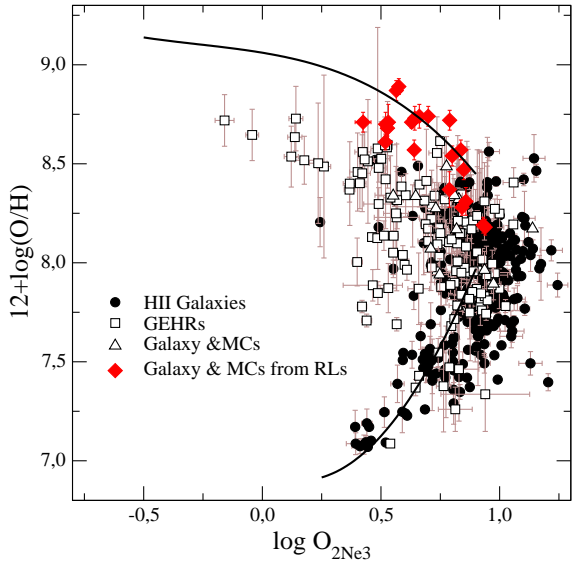}
\caption{Relation between oxygen abundance and the O$_{2Ne3}$ parameter for the sample of objects described in the text (symbols) and the calibration of McGaugh of the O$_{23}$ parameter for an average value of the ionization parameter (log([O{\sc iii}]/[O{\sc ii}]) = 1, solid line).}
\label{o2ne3}
\end{minipage}
\end{figure} 

\begin{figure}
\begin{minipage}{85mm}
\psfig{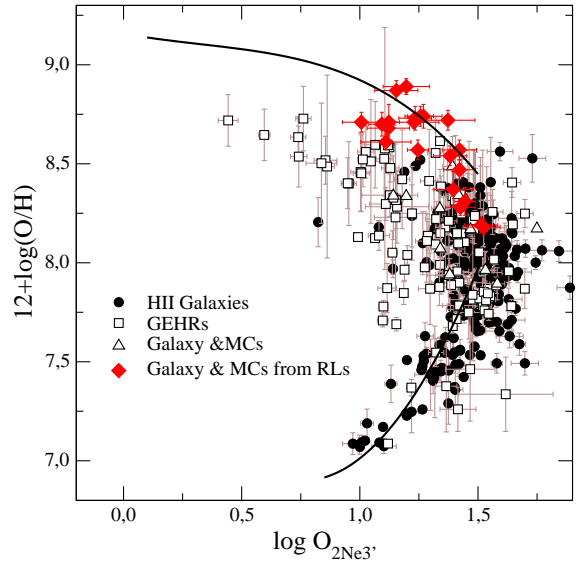}
\caption{Relation between oxygen abundance and the O$_{2Ne3'}$ parameter for the sample of objects described in the text (symbols) and the empirical calibration of O$_{23}$ of McGaugh for an average value of the ionization parameters (log([O{\sc iii}]/[O{\sc ii}]) = 1, solid line). This calibration has been displaced in the $x$ axis 0.6 dex to take into the account the mean value of the H$\beta$/H$\delta$ ratio.}
\label{o2ne3p}
\end{minipage}
\end{figure} 

\begin{figure*}
\begin{minipage}{170mm}
\centerline{
\psfig{figure=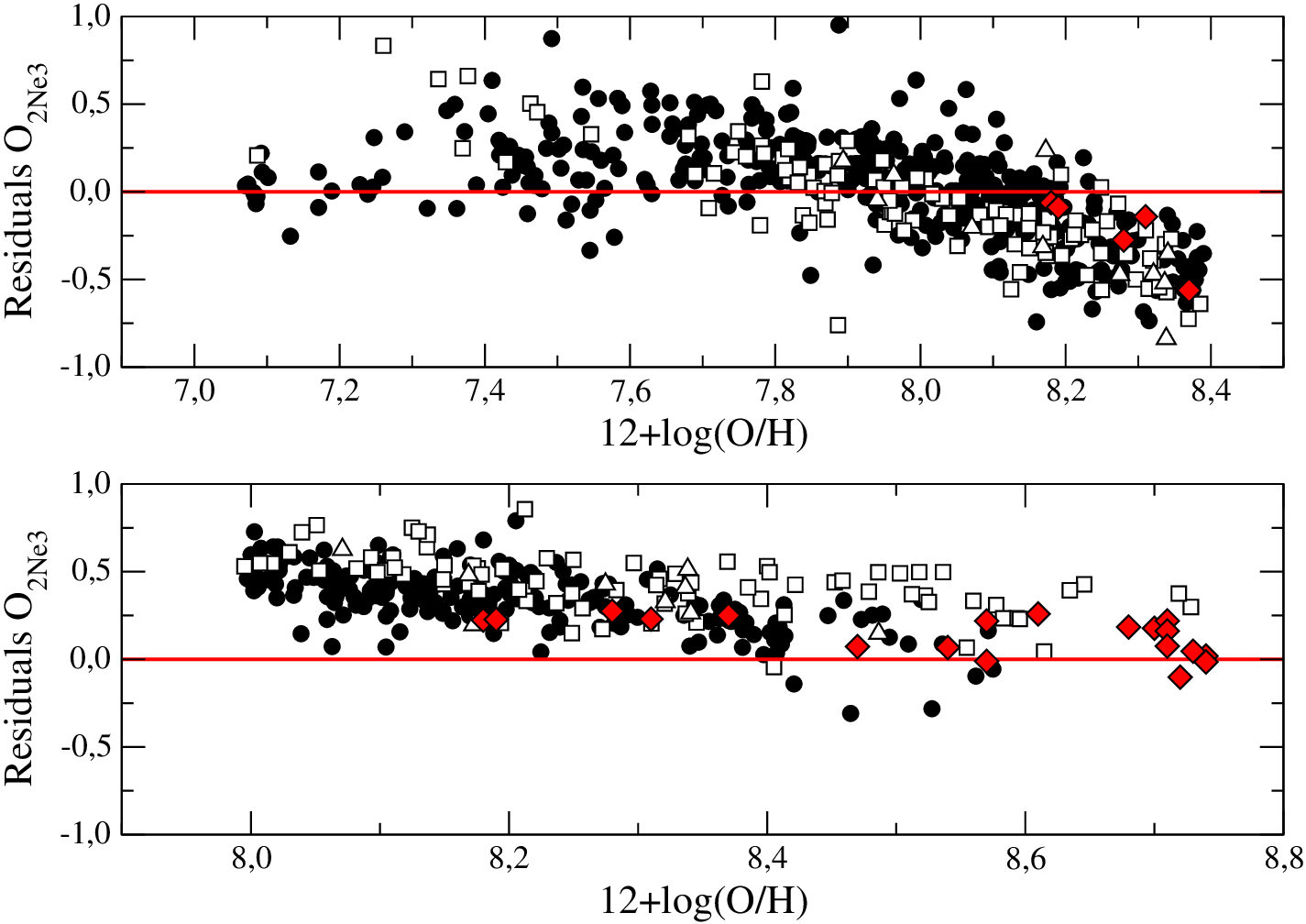,width=7.5cm,clip=}
\psfig{figure=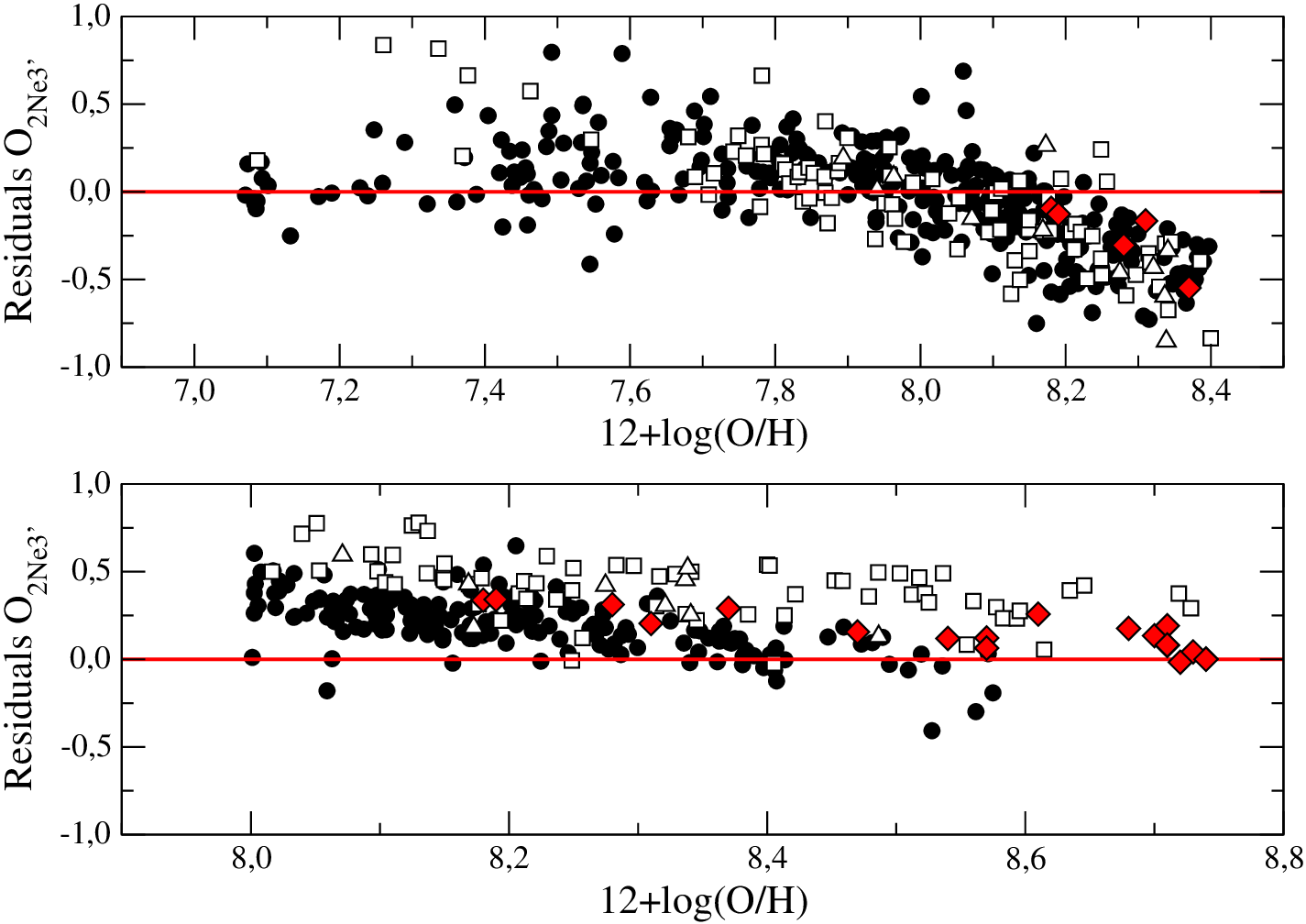,width=7.5cm,clip=}}
\caption{Residuals between the oxygen abundances derived from the McGaugh (1991) calibration of O$_{23}$ applied to O$_{2Ne3}$ (left) and to O${_2Ne3'}$  (right) and the oxygen abundances derived from the direct method. Upper panels show the calibration for low metallicites and lower panels for high metallicities.}
\label{o2ne3_res}
\end{minipage}
\end{figure*} 

We can then define a proxy for the O$_{23}$ parameter, using the relation between [O{\sc iii}] and [Ne{\sc iii}] lines, previously discussed:
\begin{equation}
O_{2Ne3} = \frac{I([O{\sc ii}] 3727\AA)+ 15.37\cdot I([Ne{\sc iii}] 3869 \AA)}{I(H\beta)} \approx O_{23}
\end{equation}
\noindent that is shown in Figure \ref{o2ne3} for the sample described in Section 2. We also show as a solid line the equivalent to the calibration of the O$_{23}$ parameter based on the models of McGaugh (1991), for an average value of the ionization parameter (log([O{\sc iii}]/[O{\sc ii}]) = 1). This calibration gives the lowest dispersion in relation with the sample of objects in both the lower and the upper branch (P\'erez-Montero \& D\'\i az, 2005). We also show the sample of objects with a direct determination of the oxygen abundances based on recombination lines in order to illustrate how this parameter has the same problems described above for O$_{23}$. We can redefine the O$_{2Ne3}$ parameter relative to the
closest and brightest hydrogen recombination line, H$\delta$. These lines are closer in wavelength, which makes its relation to be little reddening and flux calibration dependent, and useful to larger redshifts (up to z $\approx$ 1.3).
\begin{eqnarray}
O_{2Ne3'} = \frac{I([O{\sc ii}] 3727\AA)+ 15.37\cdot I([Ne{\sc iii}] 3869 \AA)}{I(H\delta)}\approx \nonumber \\
\approx O_{23} \cdot \frac{I(H\beta)}{I(H\delta)}
\end{eqnarray}

This parameter being equivalent to O$_{23}$, it suffers from its same problems, with the additional difficulty of being based on weaker lines and therefore, more difficult to measure with good signal-to-noise ratio. In the case of the hydrogen recombination line H$\delta$, it is affected more importantly by the presence of the absorption line due to underlying stellar populations and its measurement has to be carried out carefully ({\em e.g.} H\"agele et al., 2006).

\begin{figure}
\begin{minipage}{85mm}
\psfig{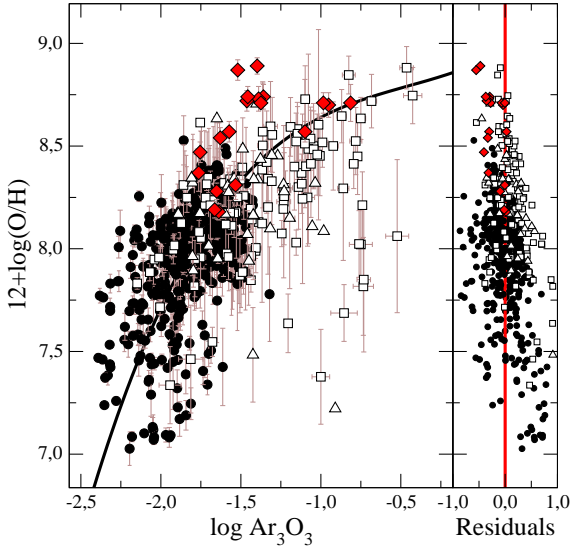}
\caption{Relation between the Ar$_3$O$_3$ parameter and the oxygen abundances derived from the direct method, along with the calibration proposed by Stasi{\'n}ska (2006) for this parameter. In the right panel, we represent the residuals between the directly derived abundances and those derived from the calibration.}
\label{ar3o3}
\end{minipage}
\end{figure} 

\begin{figure*}
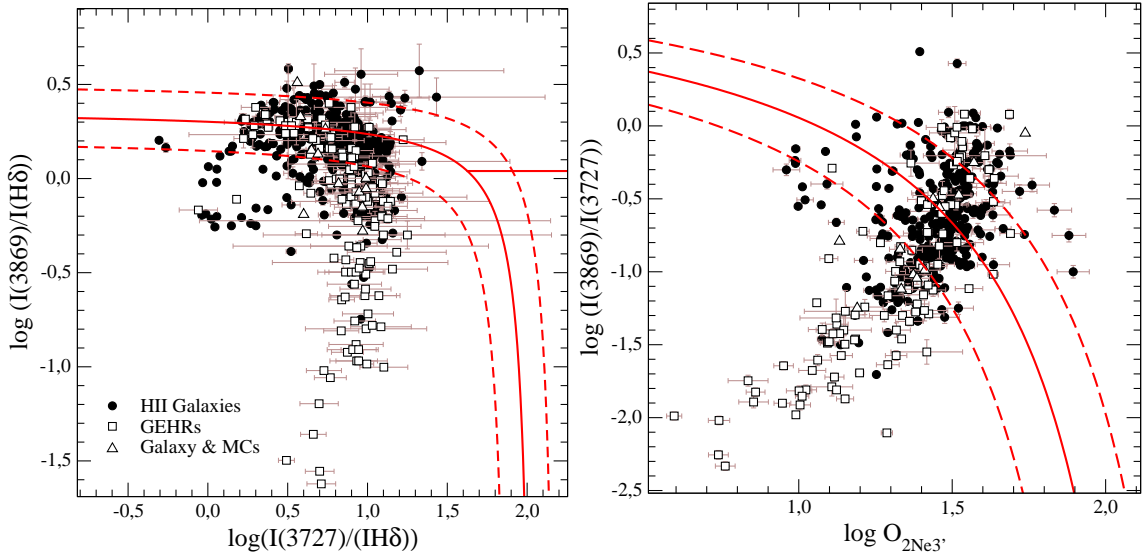

\begin{minipage}{170mm}
\centerline{
\psfig{figure=diagbluene1.eps,width=7.5cm,clip=}
\psfig{figure=diagbluene2.eps,width=7.5cm,clip=}}
\caption{Diagnostic diagrams based on neon emission lines for the sample of objects described in Section 2. At left, the relation between the ratios [Ne{\sc iii}]/H$\delta$ and [O{\sc ii}]/H$\delta$ and at right the relation between the ratio [Ne{\sc iii}]/[O{\sc ii}] and the O$_{2Ne3'}$ parameter. Solid lines represent the analytical division between star forming objects and AGNs. Dashed lines represent the limit of the bands of uncertainty 0.15 dex both sides of each relation. }
\label{diag}
\end{minipage}
\end{figure*} 

We show this parameter in Figure \ref{o2ne3p} for the sample of objects described in Section 2 and the sample of HII regions with a determination of the oxygen abundance based on recombination lines. We show as well as a solid line the empirical calibration from McGaugh but in this case we take into account the factor H$\beta$/H$\delta$ in the $x$ axis.  As in the case of O$_{23}$, it is necessary to distinguish between the upper and lower branch to derive the oxygen abundance from it. In the case of O$_{23}$, this is usually done by means of the [N{\sc ii}] lines, but they are not available in the blue part of the spectrum. Other possibility is using the [Ne{\sc iii}]/[O{\sc ii}] ratio, although it is very uncertain. From Figure \ref{o3_ne3} we can see that while objects with log([Ne{\sc iii}]/[O{\sc ii}]) $<$ -1.0 probably have high metallicities, the contrary is not true.
The residuals of the oxygen abundances derived with these new parameters as a function of the abundances measured using the direct method for both the lower and the upper branches are shown in Figure \ref{o2ne3_res}.  Since we can see, as in the case of O$_{23}$, both the upper and lower branch calibrations show large differences with
the derived metallicities in the turnover region (8.0 $<$ 12+log(O/H) $<$ 8.4).
The standard deviation of the residuals in the other ranges are quite similar to those obtained for the O$_{23}$ parameter in P\'erez-Montero \& D\'\i az (2005) for the McGaugh calibration of the upper branch (0.20 dex for O$_{2Ne3}$ and 0.18 dex for O$_{2Ne3'}$ in the  12+log(O/H) $>$ 8.4 range). The standard deviation of the residuals between the metallicites derived from the O$_{23}$ parameter are 0.06 dex for both  O$_{2Ne3}$ and O$_{2Ne3'}$ parameters. However, in the low metallicity regime, 12+log(O/H) $<$ 8.0, the dispersion of the residuals in relation with the direct method are 0.24 dex for the O$_{2Ne3}$ parameter and 0.22 dex for the O$_{2Ne3'}$ parameter, which are much higher than the 0.13 dex calculated for the O$_{23}$ parameter in P\'erez-Montero \& D\'\i az (2005). In this case the residuals between this calibration and those based on the neon emission line are 0.14 dex and 0.16 dex, respectively.

Among the other empirical calibrators based on Ne or Ar emission lines, we can analyse also the Ar$_3$O$_3$ parameter, defined by Stasi{\'n}ska (2006) as the ratio of the fluxes of the emission lines of [Ar{\sc iii}] at 7136 {\AA} and [O{\sc iii}] at 5007 {\AA}. In Figure \ref{ar3o3}, 
we show the relation between the oxygen abundances and the logarithm of this parameter for the sample of objects described in Section 2 having a measurement of the [Ar{\sc iii}] line. The standard deviation of the residuals, which are shown in the right panel of the same Figure as a function of metallicity, give a value of 0.35 dex. Besides, the behaviour of this parameter in the high metallicity regime is quite unclear due to the position of some of the objects with a direct determination of the oxygen abundance based on recombination lines, that show lower values of Ar$_3$O$_3$ than those expected for their metallicity.

\subsection{Diagnostic methods for the ionizing source in emission line objects}

Strong lines are used also in emission line-like objects to find out the nature of the main ionization mechanism: photoionization due to the absorption by the surrounding gas of UV photons emitted by massive stars, power law spectral energy distribution associated with active galactic nuclei or shock excitation. Since some of these diagnostic diagrams for the blue part of the spectrum use the emission lines of [O{\sc iii}] at 5007 and 4959 {\AA}, we could use as well the [Ne{\sc iii}] 3869 {\AA} emission line and the closest and brightest hydrogen recombination line to redefine these diagrams that would be useful up to larger redshifts and would also be less reddening dependent. This is the case of the relations proposed by Lamareille et al. (2004), as for instance the relation between ([O{\sc iii}] 5007\AA/H$\beta$) and ([O{\sc ii}] 3727\AA/H$\beta$). Using the appropriate relations, the analytical expression proposed by these authors take the following expression:
\begin{equation}
\log\left(\frac{[Ne{\sc iii}] 3869{\AA}}{H\delta}\right) = \frac{0.14}{\log ([O{\sc ii}] 3727\AA/H\delta) -2.05}+0.37
\end{equation}

In the left panel of Figure \ref{diag} we show this diagram for the objects described in Section 2. Starbursts galaxies and star forming regions lie below the solid line. This diagnostic ratio is, in fact, equivalent to the relation proposed by Rola et al. (1997) between ([Ne{\sc iii}] 3869\AA/H$\beta$) and ([O{\sc ii}] 3727\AA/H$\beta$).  The rate of coincidences between this diagnostic diagram and its predecessor based on [O{\sc iii}] lines reaches 83 \%. This diagram can be used also to separate Seyfert 2 galaxies from LINERS using the theoretical expression proposed by Lamareille et al. (2004) adapted for [Ne{\sc iii}] and H$\delta$:
\begin{equation}
\log\left(\frac{[Ne{\sc iii}]}{H\delta}\right) = 0.04
\end{equation}

All the HII galaxies which present abnormally high values of the emission line of [Ne{\sc iii}] shown in Figure \ref{o3_ne3} appear in this diagram in the Seyfert 2 zone, what is consistent with a possibly higher dependence of the  [Ne{\sc iii}] line on stellar effective temperature, as it has been previously suggested. This is supported by the fact that any of these galaxies are classified as Sy2 when other diagnostics diagrams, like [S{\sc ii}]/H$\alpha$ vs. [O{\sc iii}]/H$\beta$ or [N{\sc ii}]/H$\alpha$ vs. [O{\sc iii}]/H$\beta$ are used.

The other blue diagnostic ratio proposed by Lamareille et al. (2004) is the relation ([O{\sc iii}] 4959+5007 \AA)/([O{\sc ii}] 3727 \AA) versus O$_{23}$. The analytical expressions proposed by these authors for this diagram can be expressed in terms of [Ne{\sc iii}] and H$\delta$ adopting the form:
\begin{equation}
log\left(\frac{[Ne{\sc iii}] 3869 \AA}{[O{\sc ii}] 3727 \AA}\right) = \frac{1.5}{\log O_{2Ne3'} -2.3}+1.21
\end{equation}

The diagram is shown in the right panel of Figure \ref{diag} and, again, starbursts galaxies and star forming regions lie below this relation, which is shown as a solid line. Although the rate of coincidences between this relation and the one based on [O{\sc ii}] and [O{\sc iii}] is very high (92 \%), a larger number of objects is found  lying on the AGN region as compared to other diagnostics which somewhat questions its application.

\section{Summary and conclusions}
We have performed an analysis of a large sample of emission line objects with a direct determination of the oxygen abundance, via
the calculation of the electron temperature, and an accurate measurement of the emission lines of [Ne{\sc iii}] at 3869 {\AA} and [Ar{\sc iii}] at 7136 {\AA}. We have recalculated oxygen, neon and argon abundances taking into account the electron temperature most representative for each ionic species. The total chemical abundances have been calculated with the aid of new ionization correction factors for neon and argon based on a new grid of photoionization models computed using Cloudy v06.02 and WM-Basic stellar model atmospheres.

The new ICF for Ne yields lower abundances for low excitation objects in relation with the classical approximation O$^{2+}$/O $\approx$ Ne$^{2+}$/Ne, which does not take into account the charge transfer reaction between O$^{2+}$ and H$^0$. This new ICF agrees quite well with the fits proposed by Izotov et al. (2006) for the high and intermediate metallicity regime. Nevertheless, although we do not find any relevant dependence of this ICF on metallicity as we have not considered different X-ray sources in our models, there is some evidence of the underestimate of the ICF in the high metallicity regime. Firstly, the values found by Vermeij \& Van der Hulst (2002) for a sample of Galactic HII regions from the Ne emission lines in the mid-IR point to larger abundances of Ne$^+$ in these objects. Secondly, the study of the Ne/O ratio as a function of metallicity shows lower values of Ne/O in the high metallicity regime. The average value of this ratio agrees better with the solar abundances than the values recently obtained from X-ray observations in stellar coronae (Drake \& Testa, 2005).

Regarding Ar, we have obtained new ICFs for both Ar$^{2+}$ and Ar$^{2+}$+Ar$^{3+}$ quite similar to the values found by Izotov et al. (2006) but, again, we do not find, any relevant dependence on metallicity. According to these new ICFs, the values proposed by Izotov et al. (1994) clearly overestimate the total abundances of Ar. We propose as well new ICFs based on the ratio S$^{2+}$/(S$^{+}$+S$^{2+}$),
in order to calculate total Ar abundances using the red and far-IR wavelength range only. The study of the Ar/O ratio as a function of metallicity gives contradictory results for HII galaxies and Giants HII regions in spirals disks. For the first ones we find a constant value of the Ar/O ratio, in agreement with the expected results for the stellar production of this element. However, for GEHRs we find evidence for decreasing Ar/O with increasing metallicity. This result is found as well for the S/O ratio both in GEHRs (D\'\i az et al., 1991; P\'erez-Montero et al., 2006) and halo massive stars of our Galaxy (Israelian \& Rebolo, 2001). Taking into account that the ionization structure of Ar and S are quite similar and that they are produced in the same stellar cores it is not surprising that their ratios behave in similar ways.

We have studied some empirical parameters of metallicity based on Ne and Ar emission lines. This is the case of the [Ne{\sc iii}]/[O{\sc ii}] ratio proposed by Nagao et al. (2006). We have shown that this parameter is indeed much more sensitive to ionization parameter and effective temperature than to metallicity due to the tight relation existing between [Ne{\sc iii}] and [O{\sc iii}] emission lines. This is due to the constant value of the Ne/O ratio and the identical ionization structures of Ne$^{2+}$ and O$^{2+}$. In fact [O{\sc iii}] can be substituted by [Ne{\sc iii}]  in empirical metallicity calibrations and diagnostic diagrams. This offers the possibility of extending the range of applicability of these relations to objects with higher redshift minimizing at the same time the effects of reddening corrections and flux calibration. We have then defined an abundance parameter O$_{2Ne3'}$ equivalent to the commonly used O$_{23}$ but using the sum of the [O{\sc ii}] and [Ne{\sc iii}] lines relative to H$\delta$. Although this parameter has the same problems as O$_{23}$ (double-valued, dependence on $U$, calibration of the upper branch) it constitutes a new tool to derive oxygen abundances in large deep optical surveys of galaxies up to high redshifts ($\approx$ 1.3). Using the same principle, we have defined new diagnostic methods based only on [O{\sc ii}] and [Ne{\sc iii}], similar to those proposed by Rola et al. (1997),  that could be used in the same surveys to separate star forming galaxies from active galactic nuclei. The results obtained from these diagrams are very similar to those obtained from the other ones in the blue part of the spectrum.

\section*{Acknowledgements}

We would like to thank the referee, G.J. Ferland for many valuable suggestions and comments which have helped us to improve this paper.
This work has been supported by the CNRS-INSU (France) and its Programme National Galaxies and the project AYA-2004-08260-C03-03 of the Spanish National Plan for Astronomy and Astrophysics. Also, partial support from the Comunidad de Madrid under grant S0505/ESP/000237 (ASTROCAM) is acknoledged. AID acknowledges support from the Spanish MEC through a sabbatical grant PR2006-0049 and thanks the hospitality of the Institute of Astronomy of Cambridge. GH acknowledges support from the Spanish MEC through FPU grants and the hospitality of Observatoire Midi-Pyr\'en\'ees.

\end{document}